\documentclass{aa}
\usepackage{graphicx}
\usepackage{txfonts}

\begin{document}

\title{Further evidence for the presence of a neutron star in
\object{4U~2206+54}. {\it INTEGRAL} and VLA observations\thanks{Based on observations with {\it INTEGRAL}, an ESA project with
instruments and science data centre funded by ESA member states (especially
the PI countries: Denmark, France, Germany, Italy, Switzerland, Spain), Czech
Republic and Poland, and with the participation of Russia and the USA.}
}

\author{P. Blay\inst{1}
\and M. Rib\'o\inst{2}
\and I. Negueruela\inst{3}
\and J.~M. Torrej\'on\inst{3}
\and P. Reig\inst{4,5}
\and A. Camero\inst{1}
\and I.~F. Mirabel\inst{6}
\and V. Reglero\inst{1}}

\institute{Institut de Ci\`encia dels Materials, Universitat de Val\`encia, PO Box 22085, 46071 Valencia, Spain\\
\email{[pere.blay;ascension.camero;victor.reglero]@uv.es}
\and Service d'Astrophysique and AIM, CEA Saclay, B\^at. 709, L'Orme des Merisiers, 91191 Gif-sur-Yvette, Cedex, France\\
\email{mribo@discovery.saclay.cea.fr}
\and Departamento de F\'{\i}sica, Ingenier\'{\i}a de Sistemas y Teor\'{\i}a de la Se\~nal, Escuela Polit\'ecnica Superior, Universitat d'Alacant, Ap. 99, 03080 Alicante, Spain\\
\email{[ignacio;jmt]@dfists.ua.es}
\and IESL, Foundation for Research and Technology, 71110 Heraklion, Crete, Greece
\and University of Crete, Physics Department, PO Box 2208, 710 03
Heraklion, Crete, Greece\\
\email{pau@physics.uoc.gr}
\and European Southern Observatory, Alonso de C\'ordova 3107, Vitacura, Casilla 19001, Santiago 19, Chile\\
\email{fmirabel@eso.org}
}

\offprints{P. Blay}

\date{Received / Accepted}

\abstract{The majority of High Mass X-ray Binaries (HMXBs) behave as X-ray
pulsars, revealing that they contain a magnetised neutron star. Among the four
HMXBs not showing pulsations, and that do not show the characteristics of
accreting black holes, there is the unusual HMXB \object{4U~2206+54}. Here we
present contemporaneous high-energy and radio observations of this system
conducted with {\it INTEGRAL} and the VLA in order to unveil its nature. The
high-energy spectra show clear indications of the presence of an absorption
feature at $\sim$32~keV. This is the third high-energy observatory which
reveals marginal evidence of this feature, giving strong support to the
existence of a cyclotron resonance scattering feature, which implies a
magnetic field of 3.6$\times$10$^{12}$~G. On the other hand, the source is not
detected at centimetre radio wavelengths with a 3$\sigma$ upper limit of
0.039~mJy. The expected radio emission for an accreting black hole in the
low/hard state, inferred from X-ray flux measurements, would be at least 60
times greater than the measured upper limit. Both results firmly indicate
that, in spite of the absence of pulsations, \object{4U~2206+54} hosts a
magnetic accreting neutron star, the first one not to be observed as an X-ray
pulsar.
}

\maketitle

\section{Introduction} \label{introduction}

The vast majority of High Mass X-ray Binaries (HMXBs) harbour X-ray pulsars
(c.f. Bildsten et~al. \cite{bildsten97}), believed to be young neutron stars
with relatively strong magnetic fields ($B\sim10^{12}$~G). An important
fraction of them are wind-fed systems, in which the pulsar accretes from the
radiative wind of an OB supergiant (perhaps, in some cases, a Wolf-Rayet
star).

Among the handful of HMXBs not displaying X-ray pulsations, only three show
the typical characteristics of accreting black holes (\object{LMC~X-1},
\object{LMC~X-3} and \object{Cyg~X-1}). In four other HMXBs, pulsations have
not been discovered in spite of intensive searches, but there is no strong
evidence identifying the accreting object as a black hole. In principle, there
is no reason to attribute the lack of pulsations in all these systems to any
particular characteristic and different models have indeed been proposed to
explain some of them. There have been suggestions that
\object{2E~0236.6+6101}, whose counterpart is the B0\,Ve star
\object{LS~I~$+61^{\circ}$303}, and \object{RX~J1826.2$-$1450}, identified
with the O6.5\,V((f)) star \object{LS~5039}, may not be accreting binaries
after all, but X-ray sources powered by rotational energy from a young
non-accreting neutron star (Maraschi \& Treves \cite{maraschi81}; Martocchia
et~al. \cite{martocchia05}), although the presence of relativistic radio jets
points towards the accretion scenario (Massi et~al. \cite{massi04}; Paredes
et~al. \cite{paredes00}). In the case of \object{4U~1700$-$37}, optically
identified with the O6.5\,Iaf+ star \object{HD~153919}, a compact object of
unknown nature and mass $M_{\rm X}=2.4\pm0.3$~$M_{\odot}$ accretes material
from the wind of the massive supergiant (Clark et~al. \cite{clark02}).

The fourth HMXB not displaying pulsations is \object{4U~2206+54}, identified
with the O9p star \object{BD~$+53\degr$2790} (Negueruela \& Reig
\cite{negueruela01}; henceforth NR01). The relatively high X-ray luminosity of
\object{4U~2206+54}, $L_{\rm X}\sim 10^{35}$~erg~s$^{-1}$ (at an estimated
distance of 3~kpc; NR01), combined with its spectral shape, makes the presence
of a neutron star or a black hole in the system almost unavoidable. There are
reasons to believe that the compact object in this system is a neutron star
(NR01; Torrej\'on et~al. \cite{torrejon04}; Masetti et~al. \cite{masetti04}),
but the possibility of a black-hole has not been ruled out completely by
previous observations. Analysis of the {\it RXTE}/ASM X-ray lightcurve
revealed a 9.6~d periodicity, which is likely to be the orbital period of a
compact object (Corbet \& Peele \cite{corbet01}; Rib\'o et~al. \cite{ribo05}).
Moreover, the X-ray lightcurve displays short aperiodic variability, with
changes in the flux by a factor $\sim$10 over timescales of minutes (Saraswat
\& Apparao \cite{saraswat92}; NR01), which are typically seen in wind-fed
systems, presumably as a consequence of stochastic variability in the wind.

Interestingly, many high-energy sources not showing pulsations are
microquasars (containing either black holes or neutron stars), while pulsating
sources do not show significant radio-emission (Fender \& Hendry
\cite{fender00}). \object{4U~2206+54} shares many characteristics with the
well-known microquasar \object{LS~5039} (Paredes et~al. \cite{paredes00},
\cite{paredes02}). Both systems contain a non-supergiant late O-type star
(Clark et~al. \cite{clark01}) and a compact object that does not show
pulsations (Rib\'o et~al. \cite{ribo99}; Reig et~al. \cite{reig03}) orbiting
in a relatively close orbit when compared to the majority of HMXBs, and both
systems show evidences of wind-fed accretion (NR01; McSwain et~al.
\cite{mcswain04}) with X-ray luminosities in the range
$10^{34}$--$10^{35}$~erg~s$^{-1}$. However, an inspection of the VLA Sky
Survey (NVSS, Condon et~al. \cite{condon98}) reveals no radio emission up to a
3$\sigma$ upper limit of 1~mJy from \object{4U~2206+54}. The apparent
lack of both radio emission and pulsations does not fit within either the
typical scenario of a pulsar in a HMXB or the microquasar scenario, and a
deeper multi-wavelength approach is necessary.

In this work we present new {\it INTEGRAL} and VLA observations of the source
during the periods 2002 December--2004 September and 2003 May--June,
respectively. The possible presence of a cyclotron line, already suggested by
the data from {\it RXTE} and {\it BeppoSAX} (Torrej\'on et~al.
\cite{torrejon04}; Masetti et~al. \cite{masetti04}), and the absence of radio
emission are discussed.

\section{Observations and data analysis}

\subsection{High-energy observations}
\label{data_he}

{\it INTEGRAL} is a joint European mission in flight from 2002 October, with
three on-board high-energy instruments: the Imager on Board {\it INTEGRAL}
Spacecraft (IBIS), coupled with the {\it INTEGRAL} Soft Gamma-Ray Imager
(ISGRI) and the Pixellated Imaging Caesium Iodide Telescope (PICsIT),
sensitive to $\gamma$-rays from 15~keV up to 10~MeV; the SPectrometer in {\it
INTEGRAL} (SPI), optimised for spectroscopy in the $20$~keV--$8$~MeV energy
range; and the Joint European Monitor-X (JEM-X), consisting of twin X-ray
monitors, which provides information at lower energies (3--35~keV). An Optical
Monitoring Camera (OMC) gives source fluxes in the $V$ (550 nm) band and
complements the 3 high-energy instruments. All 4 instruments are co-aligned,
allowing simultaneous observations in a very wide energy range. A detailed
description of the mission can be found in Winkler et~al. (\cite{winkler03}). 

{\it INTEGRAL} observed the region around \object{4U~2206+54} on several
occasions during its first 22 months of Galactic Plan Survey scans
(GPSs), i.e., from 2002 December to 2004 September. We present in
Table~\ref{tab:rev_list} a summary of all the {\it INTEGRAL} revolutions
during which the source was inside the Field Of View (FOV) of ISGRI. In total,
the source was observed by ISGRI for $\sim$337~ks, but it was significantly
detected only for 27~ks\footnote{A detection is considered for ISGRI when the
detection level, which is given by the software package and is not directly
related to the number of $\sigma$ above background, lies above a value of 8.
For the typical exposure times of GPS pointings ($\sim2$~ks),  the
sensitivity limit of ISGRI lies around 2.5~count~s$^{-1}$ at 20--40~keV energy
range.}. For revolutions 70, 74, 145 and 189 only an upper limit is given
because the source had quite a marginal position in the FOV of ISGRI and the
detection is not significant enough. The JEM-X FOV is smaller and thus data
were only collected during those revolutions when {\it INTEGRAL} pointed close
enough to the source (marked in Table \ref{tab:rev_list} with the $\dag$
symbol)\footnote{A detection is considered for JEM-X when the detection level,
which is given by the software package and is not directly related to the
number of $\sigma$ above background, lies above a value 20.}. Only a few
pointings in 2003 May and June fulfill this requirement. Although SPI has the
largest FOV, it cannot acquire enough information with one exposure due to the
detector design. To achieve a S/N$\sim$10 for a source like
\object{4U~2206+54}, SPI would need around 300~ks. Nevertheless, using SPIROS
in TIMING mode (see Skinner \& Connell \cite{skinner03}) a light curve has
been attained in the 20--40~keV energy range. The obtained flux values are in
good agreement with the ISGRI data, but they have larger uncertainties
compared to them. Therefore, no data from SPI have been used in this analysis.
Data reduction has been performed with the standard Offline Analysis Software
(OSA) version 4.0, available from the {\it INTEGRAL} Science Data Centre
(ISDC)\footnote{{\tt http://isdc.unige.ch/index.cgi?Soft+download}}. A
detalied description of the software can be found in Goldwurm et~al.
(\cite{goldwurm03}), Diehl et~al. (\cite{diehl03}), Westergaard et~al.
(\cite{westergaard03}) and references therein.

\begin{table*}[]
\begin{center}
\caption{Summary of ISGRI observations of the field around \object{4U~2206+54}
during the {\it INTEGRAL} GPS. We list the revolution number, the
corresponding dates, the intervals of time when the source was in the FOV of
ISGRI during these revolutions in Modified Julian Days, the total on-source
time within these intervals, the amount of time in which the source was
detected, the mean count rate and its formal error obtained in the 20--40~keV
energy range and the  detection level value given by the software package
(where a significant detection is found for a detection level equal or above a
value of 8). For those revolutions for wich the significance of the detection
was between 1$\sigma$ and 3$\sigma$ (namely, revolutions 70, 74, 145 and 189)
we give 3$\sigma$ upper limits by scaling the detected count rates, assuming
that a 3$\sigma$ detection would correspond to a detection level of 8. The
$\dag$ symbol indicates those revolutions in which data from JEM-X allowed the
detection of the source.}
\label{tab:rev_list}
\begin{tabular}{llccccc}
\hline \hline \noalign{\smallskip}
Rev.   & Date & MJD & On-source time & Detected time & Mean count rate     & Detection level \\ 
       &      &     & (ks)           & (ks)          & (count~s$^{-1}$)    & \\              
\noalign{\smallskip} \hline \noalign{\smallskip}
~~26       & 2002 Dec 30--2003 Jan 1 & 52638.43--52640.07 &  14 & --- & ---     & ---   \\   
~~31       & 2003 Jan 14--16     & 52653.32--52655.90 &  12 & --- & ---         & ---   \\   
~~47       & 2003 Mar 03--05     & 52701.15--52701.27 & ~~8 & --- & ---         & ---   \\   
~~51       & 2003 Mar 15--17     & 52714.85--52714.96 & ~~8 & 2.2 & 4.1$\pm$0.4 & ~~9.0 \\   
~~54       & 2003 Mar 24--26     & 52722.85--52722.93 & ~~6 & 2.3 & 4.1$\pm$0.4 & ~~8.0 \\   
~~55       & 2003 Mar 27--29     & 52727.64--52727.67 & ~~4 & --- & ---         & ---   \\   
~~59       & 2003 Apr 08--10     & 52737.03--52737.10 & ~~4 & --- & ---         & ---   \\   
~~62       & 2003 Apr 17--19     & 52746.01--52746.04 &  10 & 7.3 & 3.9$\pm$0.4 & ~~8.6 \\   
~~67$\dag$ & 2003 May 01--04     & 52761.26--52762.45 & ~~8 & 6.5 & 5.9$\pm$0.5 &  12.6 \\   
~~70       & 2003 May 10--13     & 52769.93--52770.10 &  12 & --- & $<7.1$      & ~~4.1 \\   
~~74       & 2003 May 22--25     & 52781.92--52782.12 &  14 & --- & $<4.7$      & ~~2.7 \\   
~~79       & 2003 Jun 06--09     & 52796.88--52797.07 &  14 & --- & ---         & ---   \\   
~~82       & 2003 Jun 15--18     & 52805.94--52806.15 &  14 & 2.1 & 3.2$\pm$0.3 & ~~9.5 \\   
~~87$\dag$ & 2003 Jun 30--Jul 03 & 52820.92--52821.08 &  12 & 6.5 & 5.1$\pm$0.4 &  11.2 \\   
~~92       & 2003 Jul 15--18     & 52835.84--52836.00 &  12 & --- & ---         & ---   \\   
142        & 2003 Dec 12--14     & 52985.44--52985.60 &  12 & --- & ---         & ---   \\   
145        & 2003 Dec 21--23     & 52994.41--52994.62 &  12 & --- & $<7.6$      & ~~2.9 \\   
153        & 2004 Jan 14--16     & 53019.25--53019.45 &  14 & --- & ---         & ---   \\   
162        & 2004 Feb 10--12     & 53045.49--53045.69 &  14 & --- & ---         & ---   \\   
177        & 2004 Mar 26--29     & 53090.55--53091.48 &  16 & --- & ---	        & ---   \\   
181        & 2004 Apr 07--10     & 53102.88--53103.05 &  15 & --- & ---	        & ---   \\   
185        & 2004 Apr 19--22     & 53114.56--53114.70 &  13 & --- & ---	        & ---   \\   
189        & 2004 Apr 30--33     & 53126.47--53126.63 &  15 & --- & $<4.4$      & ~~4.8 \\   
193        & 2004 May 12--15     & 53138.39--53138.42 & ~~4 & --- & ---	        & ---   \\   
202        & 2004 Jun 08--11     & 53165.48--53165.67 &  15 & --- & ---	        & ---   \\   
210        & 2004 Jul 02--05     & 53189.33--53189.51 &  15 & --- & ---	        & ---   \\   
229        & 2004 Aug 29--Sep 01 & 53246.96--53247.13 &  15 & --- & ---         & ---	\\   
233        & 2004 Sep 09--12     & 53258.01--53258.80 & ~~8 & --- & ---         & ---   \\   
234        & 2004 Sep 12--15     & 53260.99--53261.92 &  17 & --- & ---         & ---   \\   
\noalign{\smallskip} \hline
\end{tabular}
\end{center}
\end{table*}

Archived {\it RXTE}/PCA lightcurves of four long observations made between
2001 October 12 and 20 have also been used. The Proportional Counter Array,
PCA, consists of five co-aligned Xenon proportional counter units with a total
effective area of $\sim$6000~cm$^{2}$ and a nominal energy range from 2~keV to
over 60~keV (Jahoda et~al. \cite{jahoda96}). In order to produce lightcurves
only the top Xenon layer in standard2 mode was used. The durations of these
observations range from 17.7 to 29.8~ks and the complete integration time
spans $\sim$100~ks. A more detailed description of these observations is given
in Torrej\'on et~al. (\cite{torrejon04}).

\subsection{Radio observations}

We observed \object{4U~2206+54} with the NRAO\footnote{The National Radio
Astronomy Observatory is a facility of the National Science Foundation
operated under cooperative agreement by Associated Universities, Inc.} Very
Large Array (VLA) at 8.4~GHz (3.6~cm wavelength) on two different epochs: 
2003 May 12 from 7:05 to 8:00 and from 11:40 to 12:52 UT (average MJD~52771.4,
during {\it INTEGRAL} revolution 70) with the VLA in its D configuration, and
2003 May 20 from 15:27 to 17:20 UT (MJD~52779.7, during {\it INTEGRAL}
revolution 73) with the VLA during the reconfiguration from D to A. The
observations were conducted devoting 10~min scans on \object{4U~2206+54},
preceded and followed by 2~min scans of the VLA phase calibrator
\object{2250+558}. The primary flux density calibrator used was
\object{1331+305} (\object{3C~286}). The data were reduced using standard
procedures within the NRAO {\sc aips} software package.

\section{Results}

\subsection{High Energies}

\subsubsection{Timing}

Analysis of the X-ray lightcurves clearly shows that the source is variable on
all timescales. However, except from the 9.6~d modulation observed in {\it
RXTE}/ASM data (see Corbet \& Peele \cite{corbet01}; Rib\'o et~al.
\cite{ribo05}) and believed to be the orbital period, no other periodic
variability has been detected so far. Unfortunately, the {\it INTEGRAL}
coverage of the source is not enough to test the presence of the orbital
periodicity. Therefore, orbital period analysis is out of the scope of this paper.

Pulse period analysis gave negative results for both our ISGRI and JEM-X
datasets. This was expected, as previous searches on similar timescales had
also failed (see NR01, Corbet \& Peele \cite{corbet01}, Torrej\'on et~al.
\cite{torrejon04} and Masetti et~al. \cite{masetti04}). 

\begin{figure*}[t!]
\center
\resizebox{0.65\hsize}{!}{\includegraphics[angle=0]{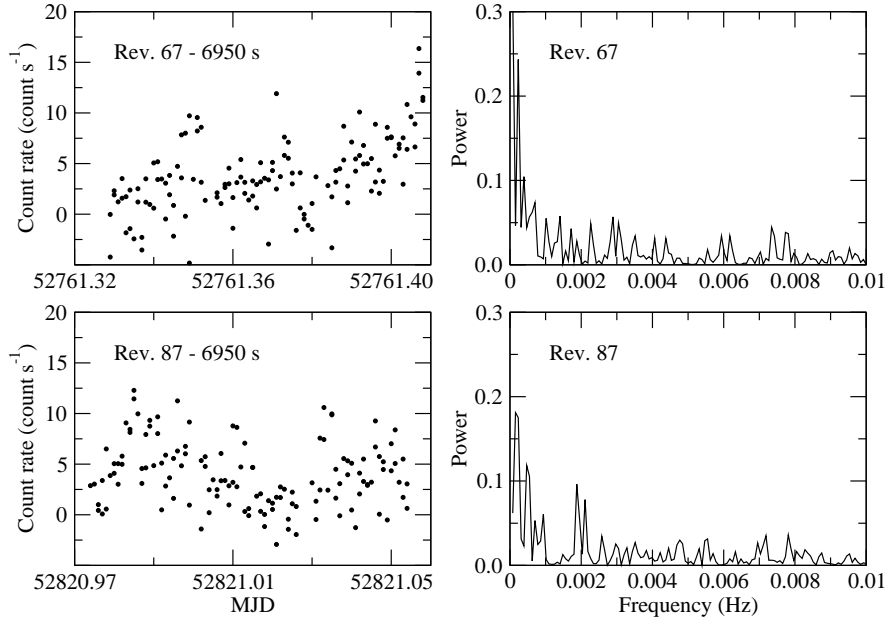}}
\caption{ISGRI lightcurves of \object{4U~2206+54} in the energy range 20--40~keV binned at 50~s (left) and associated power spectra (right) for a time span equivalent to 3 {\it INTEGRAL} science windows (around 6950~s) during revolutions 67 (top panels) and 87 (bottom panels).}
\label{fig:lc}
\end{figure*}

ISGRI data from consecutive pointings were joined together when possible and
rebinned to 50~s to search for possible longer periods. Nothing was found up
to periods of $\sim$1~h. The 20--40~keV lightcurve and power spectrum for a
time-span of 6950 seconds during revolutions 67 and 87 can be seen in
Fig.~\ref{fig:lc}. The difference between the two lightcurves and between the
corresponding power spectra is apparent. A quasi-periodic feature at
$\sim$0.002~Hz ($\sim$500~s) can be seen in data from revolution 87, but it is
not present at other epochs. The timing behaviour of the source seems to be
different in every pointing.

\begin{figure*}[t!]
\center
\resizebox{1.0\hsize}{!}{\includegraphics[angle=0]{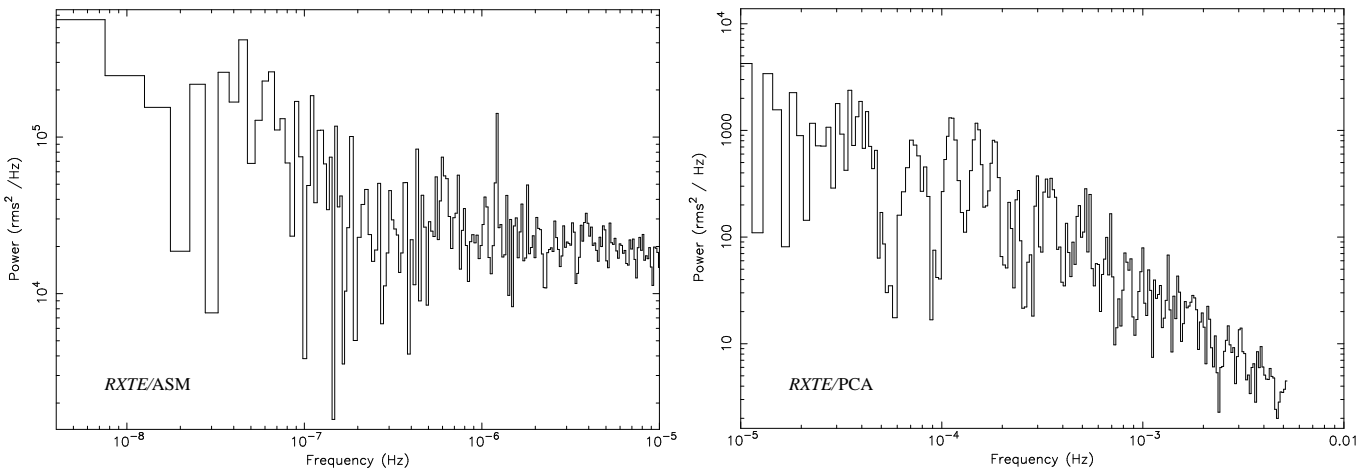}}
\caption{{\bf Left:} Power spectrum of the {\it RXTE}/ASM lighcurve, binned at
25~ks intervals. Note the feature at a frequency of
$\simeq$1.2$\times10^{-6}$~Hz, which corresponds to the orbital period (see
Corbet \& Peele \cite{corbet01}). No other period is found at the limit of the
lightcurve resolution. {\bf Right:} Power spectrum of the 100~ks {\it
RXTE}/PCA lightcurve, binned at 96~s intervals. No significant period is
detected in the frequency interval [0.95--5.5]$\times 10^{-4}$~Hz, which
corresponds approximately to the time interval from 3 hours to 30 minutes. The
slope of the power spectrum is $\sim$$\nu^{-0.8}$, where $\nu$ is the
frequency.}
\label{fig:powspec}
\end{figure*}

Little attention has been paid so far to intermediate periods (of the order of
hours), perhaps because intermediate periods would be difficult to detect in
the {\it RXTE}/ASM data, especially when points are filtered and rebinned as
one day averages to keep statistical significance. We therefore searched the
{\it RXTE}/PCA lightcurves described in Sect.~\ref{data_he} for intermediate
period pulsations to test the possible presence of a slowly rotating NS.
Unfortunately, the gaps due to the satellite low-Earth orbit are rather large
in comparison to the periods searched, which certainly hampers somewhat the
search. We used epoch folding and Lomb-Scargle periodogram techniques with
negative results. We show in Fig.~\ref{fig:powspec} the power spectrum
analysis of the whole time span. No significant period is detected,
particularly in the interval [0.95--5.5]$\times 10^{-4}$ Hz, which corresponds
to periods from 3 hours to 30 minutes, approximately.

Searching inside individual observations, we found some periodicities,
although with low significance. Inside the {\it RXTE}/PCA 60071-01-03
observation we find a possible period of 6900~s while in observation
60071-01-04 we find 8620~s. Folding the entire lightcurve on either of these
periods results in no pulsed signal. We conclude therefore that the analysis
of the entire lightcurve does not deliver any significant period in the range
explored. This result leaves only the possibility of a period in the range
from some hours to 1~d to be explored. In order to do this, a suitable long
observation with a high Earth orbit satellite, like {\it INTEGRAL}, would be
required.

\begin{figure*}[t!]
\center
\resizebox{0.65\hsize}{!}{\includegraphics[angle=0]{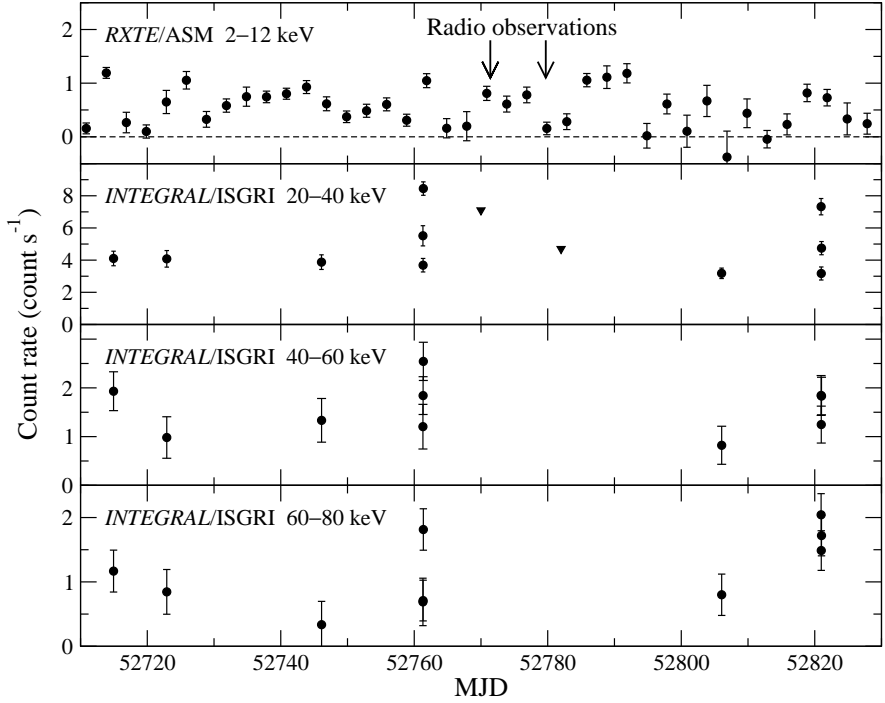}}
\caption{Lightcurves of \object{4U~2206+54} in different energy bands (quoted
inside the different panels). The top panel is for the 3-day average of the
{\it RXTE}/ASM data, where the arrows indicate the epochs of the radio
observations. The three bottom panels are for the {\it INTEGRAL}/ISGRI data in
intervals of $\sim$2~ks. Error bars represent the mean of the 1-day average
data errors for {\it RXTE}/ASM data and the formal errors for {\it
INTEGRAL}/ISGRI data. The two triangles in the 20--40~keV lightcurve indicate
3$\sigma$ upper limits.}
\label{fig:bands}
\end{figure*}

We show in Fig.~\ref{fig:bands} the long-term lightcurves of
\object{4U~2206+54} in different energy ranges, from the 2--12~keV of the {\it
RXTE}/ASM data up to 80~keV for the ISGRI data, spanning 120~d. As can be seen
in the 20--40~keV lightcurve, an increase in brightness occurred during
revolution 67 (MJD~52761.36). The source brightness increased threefold over a
timespan of the order of half an hour (see also the top left panel in
Fig.~\ref{fig:lc}).

\subsubsection{Spectral Analysis}

From the whole timespan during which ISGRI collected data from
\object{4U~2206+54}, the source was inside both the Fully Coded Field Of View
(FCFOV) of ISGRI and the JEM-X FOV only during one pointing in revolution 67
and 3 pointings in revolution 87. Therefore, the available spectrum from
revolution 67 and the mean spectrum from revolution 87 were used for the
spectral analysis. Systematic errors of 10\% for ISGRI and 5\% for JEM-X were
added to our data sets in order to perform a more realistic spectral
analysis\footnote{Private communication from the teams of the instruments.}.
The software package used was {\sc xspec} 11.2 (Arnaud \cite{arnaud96}).

With the aim of comparing with previously published data, the comptonisation
model of Sunyaev \& Titarchuk (\cite{sunyaev80}), improved by Titarchuk
(\cite{titarchuk94}) including relativistic effects, implemented in {\sc
xspec} as {\tt compTT}, and a powerlaw model, modified to include photon
absorption and a high energy cut-off, were chosen to fit the data. For the
comptonisation model, the emitting region temperatures derived were 10$\pm$3
and 13$\pm$8~keV for the data of revolutions 67 and 87, respectively. The fits
were acceptable, with corresponding $\chi^2_{\rm Red}$ of 1.3 for 173 degrees
of freedom (DOF) in the first observation, and $\chi^2_{\rm Red}$ of 1.2 for
176 DOF in the second observation. The powerlaw parameters of both
observations are listed in Table~\ref{table:comparison}.

\begin{table*}[t!]
\begin{center}
\caption[]{Comparison of most recently published powerlaw+high-energy cutoff
parameters, adding photon absorption at lower energies, for
\object{4U~2206+54}. We remit the reader also to tables 2, 3, and 4 of
Saraswat \& Apparao (\cite{saraswat92}) for earlier {\it EXOSAT} data. All
fluxes shown are unabsorbed in the quoted energy ranges. The errors are at a
68\% confidence level for the entry of Negueruela \& Reig
(\cite{negueruela01}) and at 90\% confidence level for the rest of the data
except for the entry of Corbet \& Peele (\cite{corbet01}), where confidence
levels were not reported.}
\label{table:comparison}
\begin{tabular}{@{}l@{~~}c@{~~~}c@{~~~}c@{~~}c@{~~~~}l@{}r@{~~~~}l@{}c@{}}
\hline \hline \noalign{\smallskip}
Ref. (Mission, year) & $\Gamma$ & $E_{\rm cut}$ & $E_{\rm fold}$ & $N_{\rm H} \times10^{22}$ & ~~~~$\chi^{2}_{\rm Red}$ & Flux $\times$10$^{-10}$  & Energy range & $E_{\rm cycl}$\\
                     &          & (keV)         & (keV)          & (atom~cm$^{-2}$)          & ~~(DOF)                  & (erg~s$^{-1}$~cm$^{-2}$) & (keV)        & (keV)\\
\noalign{\smallskip} \hline \noalign{\smallskip}
Negueruela \& Reig \cite{negueruela01} ({\it RXTE}, 1997) & 1.7$\pm$0.3         & 7.4$\pm$0.2 & 17.5$\pm$0.8   & 4.7$\pm$0.2  & 0.9(56)   &  4.8 & 2.5--30 & not reported \\
Corbet \& Peele \cite{corbet01}    ({\it RXTE}, 1997-1)   & 1.71$\pm$0.03       & 7.3$\pm$0.1 & 17.3$\pm$0.6   & 4.6$\pm$0.2  & 0.82$^a$  & 3.1 & 2--10 & not reported\\
~~~~~~~~~~~~~~~~~~~~~~~~~~~~~~~~~~~({\it RXTE}, 1997-2)   & 1.12$\pm$0.12       & 5.3$\pm$0.2 & 10.5$\pm$1.2   & 2.7$\pm$0.7  & 0.75$^a$  &  1.1 & 2--10   & not reported \\
Torrej\'on et~al. \cite{torrejon04} ({\it RXTE}, 1997)    & 1.6$\pm$0.1         & 7.6$\pm$0.4 & 16.3$\pm$1.2   & 4.5$\pm$0.4  & 0.71(52)  &  2.7 & 2--10   & not reported\\
~~~~~~~~~~~~~~~~~~~~~~~~~~~~~~~~~({\it RXTE}, 2001)       & 1.6$\pm$0.1         & 4.3$\pm$0.3 & 20$\pm$2       & 4.6$\pm$0.1  & 1.27(49)  &  1.3 & 2--10   & 29$^{b}$\\
~~~~~~~~~~~~~~~~~~~~~~~~~~~~~~~~~({\it BeppoSAX}, 1998)   & 1.0$\pm$0.2         & 7.8$\pm$0.5 & 11$\pm$3       & 1.1$\pm$0.3  & 1.32(113) &  0.4 & 2--10   & ~35$\pm$5$^{b}$\\
Masetti et~al. \cite{masetti04} ({\it BeppoSAX}, 1998)    & 0.95$^{+0.11}_{-0.14}$ & 4.3$^{+0.6}_{-0.5}$ & 10.6$^{+2.7}_{-2.0}$ & 0.88$^{+0.21}_{-0.19}$ & 1.1(219) & 0.4 & 2--10 & ~35$\pm$5$^{c}$\\
This paper ({\it INTEGRAL}, 2003, Rev. 67)                & 1.8$\pm$0.7         & 13$\pm$5    & 22$\pm$6       & 1.0\,(fixed) & 1.2(154)  & 15.9 & 4--150  & 32$\pm$5 \\
This paper ({\it INTEGRAL}, 2003, Rev. 87)                & 1.7$^{+0.3}_{-0.4}$ & 11$\pm$5    & 29$^{+8}_{-7}$ & 1.0\,(fixed) & 1.0(153)  &  8.3 & 4--150  & 32$\pm$3\\
\noalign{\smallskip} \hline
\end{tabular}
\end{center}
$^{a}$ No information about the DOF is reported in the reference.
$^{b}$ No significance reported.
$^{c}$ At 2$\sigma$ confidence level.
\end{table*}

Both models yield a 4--150~keV flux of $\sim$16$\times$10$^{-10}$
erg~s$^{-1}$~cm$^{-2}$ for revolution 67 and $\sim$8$\times$10$^{-10}$
erg~s$^{-1}$~cm$^{-2}$ for revolution 87. Assuming a distance to the source of
3~kpc (NR01), its luminosity amounts to $\sim$1.7$\times$10$^{36}$
erg~s$^{-1}$ and $\sim$0.9$\times$10$^{36}$ erg~s$^{-1}$, respectively. Around
$\sim$50\% of the total luminosity lies in the 4--12~keV energy band, that is
$\sim$8.5$\times$10$^{35}$ erg~s$^{-1}$ for revolution 67 and
$\sim$4.5$\times$10$^{35}$ erg~s$^{-1}$ for revolution 87. We notice that
during these observations the source appears brighter than in any previous
observation. The {\it RXTE}/ASM lightcurve confirms that the flux was high in
the 2--12~keV band as well.

\begin{figure}[t!]
\center
\resizebox{1.0\hsize}{!}{\includegraphics[angle=0]{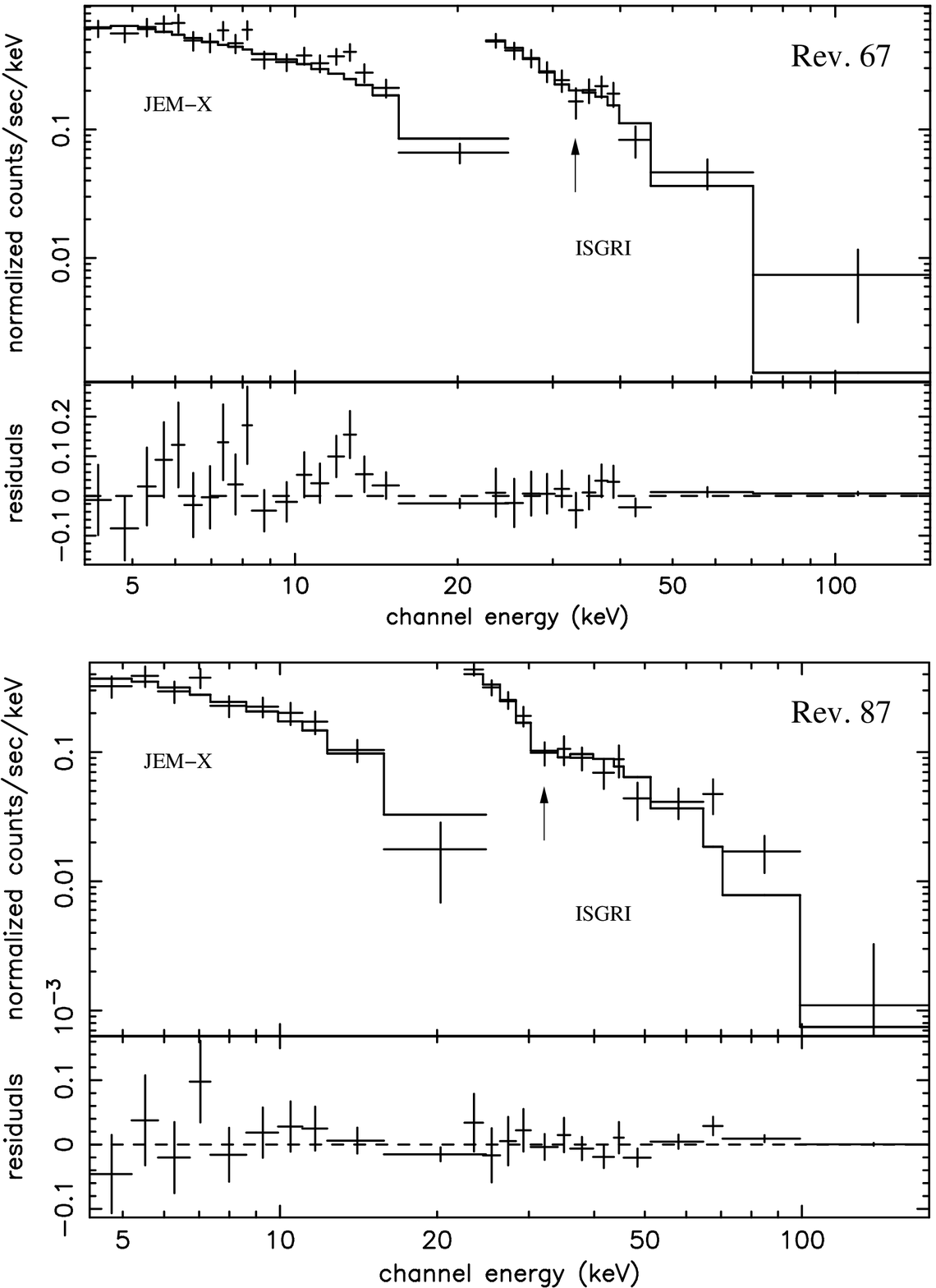}}
\caption{Spectra of \object{4U~2206+54} from JEM-X and ISGRI detectors. {\bf
Top:} The spectrum shown is for 2.2~ks of exposure time during revolution 67
(MJD~52761.36). {\bf Bottom:} The spectrum shown is for 6.6~ks of exposure
time during revolution 87 (MJD~52821.17). In both cases the spectral model
shown in Table~\ref{table:comparison} is represented by the solid line, and
the residuals to the model are displayed in the lower panels. The presence of
an absorption feature around 32~keV, indicated by arrows, is suggested by the
two datasets.}
\label{fig:spe_r67_r87}
\end{figure}

We show in Fig.~\ref{fig:spe_r67_r87} the {\it INTEGRAL} spectra of
\object{4U~2206+54} for revolutions 67 and 87. Both spectra suggest the
presence of an absorption feature around $\sim$30~keV, as already noticed by
Torrej\'on et~al. (\cite{torrejon04}) and Masetti et~al. (\cite{masetti04}) in
{\it RXTE} and {\it BeppoSAX} data, see Table~\ref{table:comparison}. An
absorption feature through the {\tt cyclabs} model (in {\sc xspec} notation)
was added to the powerlaw model. In the revolution 67 spectrum, the absorption
feature was fitted at an energy of 32$\pm$5~keV for a fixed line width of
3~keV. The same feature is apparent in the spectrum from revolution 87, where
it was fitted at 32$\pm$3~keV \footnote{An F-test was applied to the spectral fits
of revolution 87, wich has a better signal to noise ratio than that of revolution 67.
The improvement of the $\chi^2$ by the inclusion of a cyclabs component has a 12\%~
probability of ocurring by chance. One 
should take into account the limitations of this test when applied to lines 
(Protassov et al. \cite{protassov02})}. Except for the normalization factors, the
fitted parameters to the datasets of both revolutions are compatible between
them within the errors (see Table \ref{table:comparison}).



\begin{figure}[t!]
\center
\resizebox{1.0\hsize}{!}{\includegraphics[angle=0]{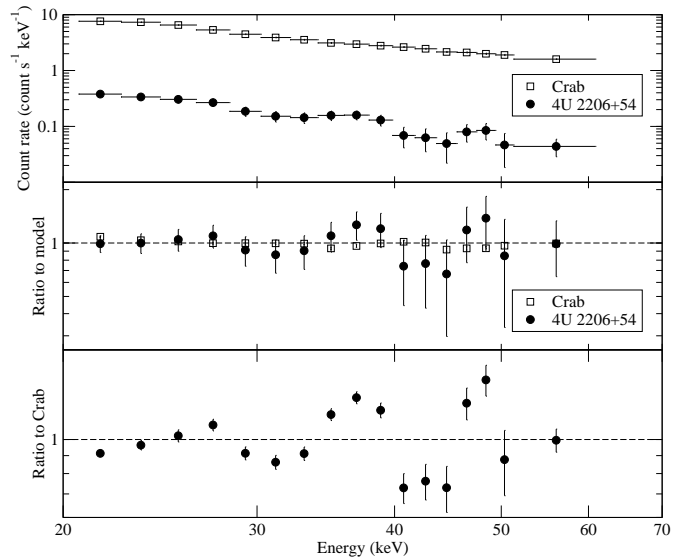}}
\caption{Observed ISGRI spectra of the \object{Crab} and \object{4U~2206+54}
during revolution 67 (top panel), their ratio to the powerlaw model (middle
panel), and the ratio of \object{4U~2206+54} to that of the \object{Crab} for
the former ratios. The variations above 40~keV are not smooth, but random, and
show large errors, indicating that they reflect most likely extraction
problems due to the fact that we are approaching the sensitivity limit of
ISGRI ($\sim$0.04~count~s$^{-1}$~keV$^{-1}$ for 2~ks exposures and 3$\sigma$
level detection at 45~keV).}
\label{fig:ratio_to_crab}
\end{figure}

There are well known calibration problems in the ISGRI Response Matrix
Function (RMF) that may cast some doubts about the reality of the absorption
feature reported here. In order to investigate if this feature is an
instrumental effect, we normalised the 20--60~keV spectra of
\object{4U~2206+54} and \object{Crab} to their respective continua modeled by
a powerlaw, and then divided the normalised \object{4U~2206+54} spectrum over
that of the \object{Crab}. We have chosen a \object{Crab} observation as close
in time as possible to our data and with similar off-axis angles, to ensure
that the RMF and off-axis effects are as much similar as possible. We show in
Fig.~\ref{fig:ratio_to_crab} the observed spectra (top panel), their ratio to
the powerlaw model (middle panel), and the ratio between the former ratios
(bottom panel). The absorption feature around 32~keV is still seen. The
quality of the data does not allow us to state that the detection is
statistically significant, but the likely presence of a feature at this
position had already been reported in the analysis of two other independent
datasets obtained by two different satellites (see Table
\ref{table:comparison}). As it has been seen by three different instruments at
different times, the existence of this absorption feature is strongly
suggested. Such features in X-ray spectra are generally attributed to
Cyclotron Resonance Scattering Features (CRSFs) (see, e.g., Coburn et~al.
\cite{coburn02} and references therein).

\begin{figure}[t!]
\center
\resizebox{1.0\hsize}{!}{\includegraphics[angle=0]{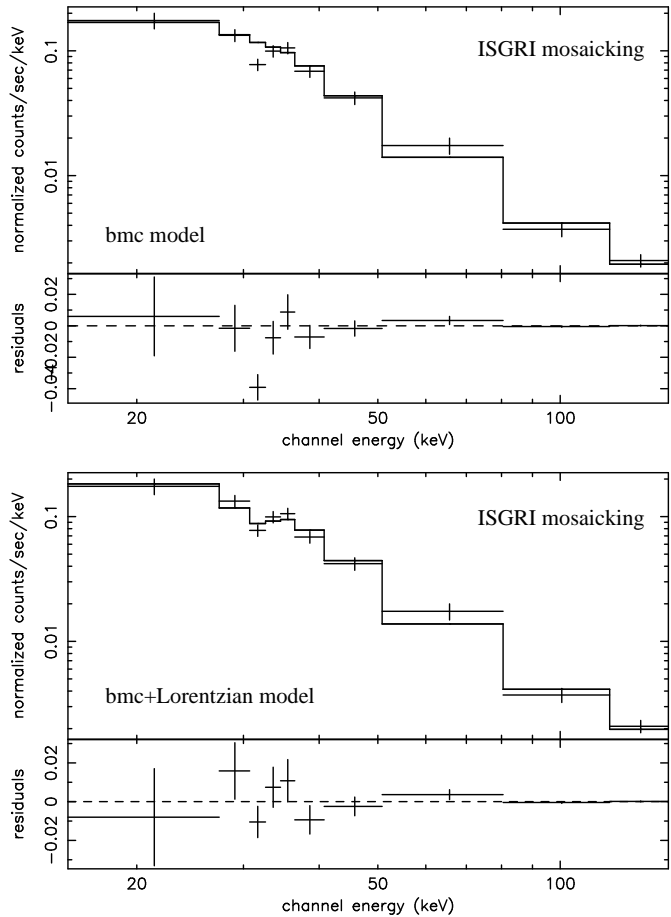}}
\caption{{\bf Top:} Spectrum of \object{4U~2206+54} extracted from the
mosaicking of all images with significant detections from ISGRI. The solid
line represents the {\tt bmc} model spectrum (in {\sc xspec} notation) fitted
to the data. An absorption feature around 32~keV is seen in the residuals.
{\bf Bottom:} The same extracted spectrum, but here the solid line represents
the {\tt bmc} model plus a Lorentzian feature in absorption fitted to the
data, yielding a line position of 31.5$\pm$0.5~keV.}
\label{fig:mosa_spe}
\end{figure}

Motivated by the possible presence of this CRSF, we have summed up images from
those {\it INTEGRAL} revolutions with significant ISGRI detections. The
effective exposure time of this mosaic amounts to $\sim$27~ks, and we show its
extracted spectrum in Fig.~\ref{fig:mosa_spe}. We note that spectral shape
changes with luminosity have been reported in Saraswat \& Apparao
(\cite{saraswat92}) and NR01. Therefore, by summing up data taken on different
epochs we might be losing spectral shape information. However, our goal is not
to study the shape of the continuum, but to achieve an improved signal to
noise ratio at the CSRF position, which is suggested by both the revolution 67
and revolution 87 spectra to be at $\sim$32~keV. The best fit to the continuum
of the new spectrum was a comptonisation model of soft photons by matter
undergoing relativistic bulk-motion, i.e. {\tt bmc} in {\sc xspec} notation
(Shrader \& Titarchuk \cite{shrader99}), which provided a $\chi^{\rm 2}_{\rm
Red}$ of 1.7 for 6 degrees of freedom (see top panel in
Fig.~\ref{fig:mosa_spe}). We added a CSRF absorption feature, by using
the {\sc xspec} {\tt cyclabs} model, at 32$\pm$2~keV to the {\tt bmc} model,
and obtained a slightly improved fit with $\chi^{\rm 2}_{\rm Red}$ of 1.3 for
5 degrees of freedom. Finally, we also fitted the data by adding a Lorentzian
profile in absorption to the {\tt bmc} model. In this case the $\chi^{\rm
2}_{\rm Red}$ lowered to 1.1 for 5 degrees of freedom (see bottom panel in
Fig.~\ref{fig:mosa_spe}). The center of the line was located at
31.5$\pm$0.5~keV and its FWHM was found to be 0.015$\pm$0.005~keV. The
normalization of the line, 1.5$^{+0.7}_{-0.8}\times$10$^{-3}$
count~s$^{-1}$~keV$^{-1}$ at a 68\% confidence level (1$\sigma$), yields a
significance of the line of $\sim$2$\sigma$\footnote{The
significance of the CRSF detection could not be derived from the individual analysis 
of revolutions 67 and 87, due to the poor improvement of the statistics obtained when adding
the CRSF component in both cases. We can only state the significance of the detection
after the nice improvement in the signal-to-noise ratio
achieved by the mosaic.}.  We note that the
significance of this detection might be somehow influenced by changes in the
continuum during the different observations. On the other hand, although the
significance depends as well on the model chosen to fit the continuum, any
model that properly fits the obtained continuum will reveal the presence of a
$\sim$2$\sigma$ absorption around 32~keV, as can be seen from the data of Fig.
\ref{fig:mosa_spe}. Thus we can conclude that the presence of a CRSF is
strongly suggested by the data. This result, when combined with the previous
claims after {\it BeppoSAX} and {\it RXTE} data (Torrej\'on et~al.
\cite{torrejon04}; Masetti et~al. \cite{masetti04}) gives evidence for the
presence of this CSRF in \object{4U~2206+54}.

\subsection{Radio} \label{radio}

\begin{figure}[t!]
\center
\resizebox{1.0\hsize}{!}{\includegraphics{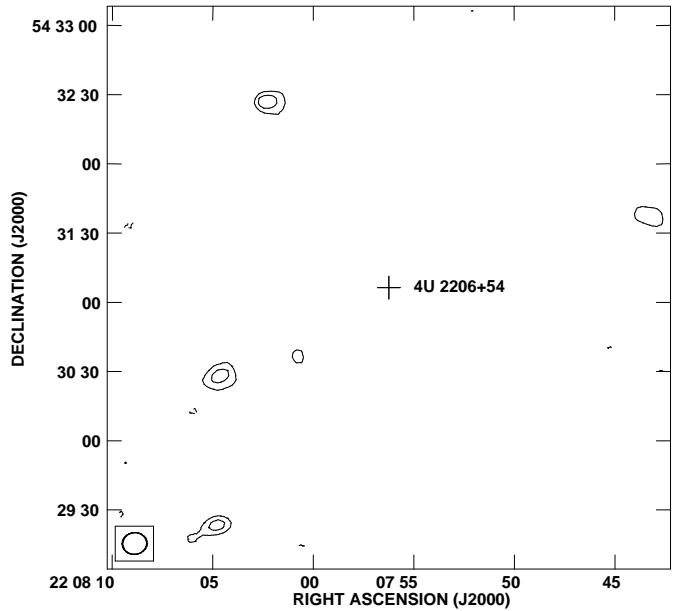}}
\caption{Image around \object{4U~2206+54}, marked with a cross, obtained with
the VLA at 8.4~GHz after concatenating data from 2003 May 12 and 20. The image
size is 4\arcmin$\times$4\arcmin. Contours are $-$3, 3, and 5 times the rms
noise level of 0.013~mJy~beam$^{-1}$. The ellipse in the bottom left corner
represents the Full Width Half Maximum of the obtained synthesised beam of
10.6\arcsec$\times$9.4\arcsec\ in PA=$-$83.4\degr.}
\label{fig:vla}
\end{figure}

No radio emission at 8.4~GHz was detected, with a 3$\sigma$ upper limit of
0.042~mJy on 2003 May 12 (MJD~52771.4) and a 3$\sigma$ upper limit of
0.066~mJy on 2003 May 20 (MJD~52779.7). We concatenated all the data and
obtained a final 3$\sigma$ upper limit of 0.039~mJy. The resulting image is
shown in Fig.~\ref{fig:vla}.

\section{Discussion} \label{discussion}

\subsection{Conflicts in the neutron star scenario}

The aperiodic variability of the X-ray emission from \object{4U~2206+54}
favours the idea that the X-ray source is powered by wind-fed accretion onto a
compact object. The X-ray luminosity of the source combined with its X-ray
spectral shape likely excludes the possibility of a white dwarf (see, e.g., de
Martino et~al. \cite{demartino04} for accreting white dwarf X-ray spectra). The compact
object must be, therefore, a neutron star or black hole.

There are two main difficulties for accepting \object{4U~2206+54} as a typical
wind-accreting neutron star. The first one is the lack of pulsations, as most
other wind-fed systems are X-ray pulsars. In principle, this might result from
a geometrical effect: if the angle between the spin axis and the magnetic axis
of the neutron star is close to zero or the whole system has a very low
inclination angle, all the high-energy radiation seen could be coming from a
permanently observed single pole of the neutron star. The system inclination
is unlikely to be very small, as the projected $v\sin i$ for the optical
companion is not particularly small (NR01), unless there is a very strong
misalignment between the rotation axis of the optical star and the orbit.
However, if the angles between the spin and magnetic axes of neutron stars are
drawn from a random distribution, there is a non-negligible chance that for
some systems they will be aligned. Similar scenarios have been proposed to
explain the absence of pulsations from \object{4U~1700$-$37} (White et~al.
\cite{white83}) and also from the low-mass X-ray binary \object{4U~1700+24}
(Masetti et~al. \cite{masetti02}), though in the latter case there is no
conclusive evidence that this system is sufficiently young to show pulsations.

The second, stronger argument is the expected X-ray luminosity. We can
consider a canonical neutron star with 10~km radius and 1.4~$M_\odot$
accreting from the fast wind of a low-luminosity O9\,III--V star in a 9.6~d
orbit. Following the Bondi-Hoyle approximation, the accretion luminosity
obtained, which is an upper limit to the X-ray luminosity, is
$\sim$1$\times$10$^{34}$~erg~s$^{-1}$ (see Reig et~al. \cite{reig03} for
details about the method). In contrast, the observed X-ray luminosity of
\object{4U~2206+54} is in the range $\sim$10$^{35}$--10$^{36}$~erg~s$^{-1}$,
therefore comparable to those of HMXBs with OB supergiant donors (see
Negueruela \cite{negueruela04} and references therein), which are believed to
have mass-loss rates more than one order of magnitude higher than a O9\,III--V
star (Leitherer \cite{leitherer88}; Howarth \& Prinja \cite{howarth89}). In
addition, we note that our estimate for the semimajor axis of
\object{4U~2206+54}, of 55--60~$R_\odot$, is comparable to the highest values
of semimajor axes in supergiant systems (see Kaper et~al. \cite{kaper04}).
Therefore, a close orbit cannot be invoked to solve the problem of the high
X-ray luminosity.

\subsection{Excluding the black hole scenario}

If a black hole was present in the system, the photon index and luminosities
of our {\it INTEGRAL} observations would indicate that the source is in a
low/hard state (see McClintock \& Remillard \cite{mcclintock04} and references
therein). Our radio observations took place on 2003 May 12 and 20, or during
{\it INTEGRAL} revolutions 70 and 73, i.e., right between {\it INTEGRAL}
observations at revolutions 67 (2003 May 01--04) and 87 (2003 Jun 30--Jul 03).

Gallo et~al. (\cite{gallo03}) found an empirical correlation between the soft
X-ray flux (in the range 2--11~keV) and the centimetre radio emission (with 
observed flat spectrum in the range 4.9--15~GHz) for black hole binary
systems in the low/hard state, of the form: $S_{\rm radio}=(223\pm156)\times
(S_{\rm X})^{+0.7}$, where $S_{\rm radio}$ is the radio flux density scaled to
1~kpc, $S_{\rm X}$ is the X-ray flux in Crab units scaled to 1~kpc, and the
uncertainty in the multiplying factor is the non-linear 1$\sigma$ error of
their fit. Therefore, by using a measured X-ray flux we can compute the
expected radio emission of a source in case it is a black hole.

We obtained the flux from \object{4U~2206+54} in the 2--11~keV band from our
JEM-X data, being 7.2 and 4.0$\times10^{-10}$ erg~s$^{-1}$~cm$^{-2}$ for the
2003 May and June observations, respectively. This flux was translated to
\object{Crab} units by measuring the flux from the \object{Crab} in the
2--11~keV band, using JEM-X data from an {\it INTEGRAL} observation close in
time to our \object{4U~2206+54} pointings. The \object{Crab} flux was found to
be $1.8\times10^{-8}$~erg~s$^{-1}$~cm$^{-2}$, leading to fluxes of 40 and
22~mCrab for \object{4U~2206+54} during revolutions 67 and 87, respectively.
From this, and using $N_{\rm H}= 1.0\times10^{22}$~atom~cm$^{-2}$ (average of
the values obtained by Torrej\'on et~al. \cite{torrejon04} and Masetti et~al.
\cite{masetti04} from {\it BeppoSAX} data) we computed the unabsorbed
corrected flux following equation (1) of Gallo et~al. (\cite{gallo03}), and
then the resulting flux in Crab units scaled to 1~kpc distance (assuming a
distance of 3~kpc to \object{4U~2206+54}). The relation discussed above then
predicts a radio flux density, already scaled back again to 3~kpc, of
12.6$\pm$8.8~mJy at the time of revolution 67 and of 8.3$\pm$5.8~mJy for
revolution 87 (where the errors come directly from the 1$\sigma$ uncertainties
given in Gallo et~al. \cite{gallo03} for the parameters of their fit). Thus,
for revolution 67 the expected radio emission would be in the range
3.8--21.4~mJy, and for revolution 87 in the range 2.5--14.2~mJy.

We note that the lower expected radio flux density of 2.5~mJy is already more
than 60 times greater than the 0.039~mJy 3$\sigma$ upper limit found with our
VLA observations. Obviously, it can be argued that our observations were not
simultaneous. During revolution 70 (which was in coincidence with the first
radio observations) the flux found from ISGRI data is not significant enough
and, unfortunately, the source was outside the FOV of JEM-X, which could have
provided an X-ray flux suitable for this analysis. Nevertheless, we point out
that the {\it RXTE}/ASM count rate during our first radio observation is very
similar to that measured during revolution 87, so it is reasonable to compare
the obtained 2.5~mJy limit with our measured 0.042~mJy 3$\sigma$ upper limit
on that day, giving again a difference of a factor $\sim$60. It could also
be argued that the source could have experienced a transition to the high/soft
state, that would naturally prevent the detection of radio emission. However,
in such a case the {\it RXTE}/ASM count rates should have increased
considerably, while during both radio observations the count rates were
similar or lower than during revolutions 67 and 87, when the photon indexes
were typical of low/hard states. In summary, if the correlation between X-ray
emission and radio emission reflects indeed a general property of black hole
systems, we conclude that there is not a black hole in \object{4U~2206+54}.

Moreover, Fender \& Hendry (\cite{fender00}) show that all Galactic persistent
black holes have detectable radio emission. As \object{4U~2206+54} is a
persistent source and does not show any detectable radio emission, it cannot
host a black hole. Systems containing magnetised neutron stars
($B\gtrsim10^{11}$~G), on the other hand, do not show detectable radio
emission. 

\subsection{The cyclotron feature}

The indication of a cyclotron feature centred at 32~keV strongly suggests that
there is a magnetic neutron star in \object{4U~2206+54}, in good agreement
with the lack of radio emission. {\it INTEGRAL} is the third mission reporting
the likely detection of this absorption feature (see Table
\ref{table:comparison}) and, even if none of the detections can be considered
statistically significant, the fact that it appears in three independent
datasets cannot be ignored.

If the line is indeed a CRSF we can compute the value of the magnetic field in
the scattering region by means of the equation $[B/10^{12}~{\rm G}]=[E_{\rm
cycl}/11.6~{\rm keV}]\,(1+z)$, being $z$ the gravitational redshift at which
we see the region. Considering that the line is produced at the surface of a
canonical neutron star of 1.4~$M_\odot$ with a radius of 10~km, the
gravitational redshift amounts to $z$=0.3 (see, e.g., Kreykenbohm et~al.
\cite{kreykenbohm04}), and from the position of the line centre at 32~keV we
obtain a magnetic field of $3.6\times10^{12}$~G. This value, in agreement with
those found by Torrej\'on et~al. (\cite{torrejon04}) and Masetti et~al.
(\cite{masetti04}), is typical of magnetic neutron stars, and well within the
range of $1.3$--$4.8\times10^{12}$~G obtained by Coburn et~al.
(\cite{coburn02}) for a sample of ten X-ray pulsars displaying CRSFs (see
their table~7).

One is led to the conclusion that \object{4U~2206+54} is the first system
known in which an accreting magnetic neutron star does not appear as an X-ray
pulsar. In principle, the possibility of very slow pulsations cannot be
discarded. The wind-accreting X-ray source \object{2S~0114+650}, with a B1
supergiant donor (Reig et~al. \cite{reig96}), shows pulsations with a period
of $\sim$2.8~h (Finley et~al. \cite{finley94}). The orbital period of the
system is $\sim$12~d, similar to that of \object{4U~2206+54}. Our timing
analysis rules out the possibility of significant pulsations from
\object{4U~2206+54} in the range 0.5--3.0~h. A modulation at a period of
several hours is still possible, as existing datasets do not constrain
strongly this period range. However, it seems more logical to conclude that
geometrical effects are responsible for the lack of pulsations.

\section{Conclusions} \label{conclusions}

We present the first {\it INTEGRAL} GPS results on \object{4U~2206+54}
together with contemporaneous VLA observations of the source. A broad
high-energy spectrum (4--150~keV), joining JEM-X and ISGRI data, has been
extracted and fitted with spectral models similar to those used in the
analysis of previously published data obtained with other satellites. The
evidence for the presence of a cyclotron line is furthered, as {\it INTEGRAL}
becomes the third high-energy mission to observe a possible feature at
32$\pm$3~keV. If the feature is indeed a cyclotron line, it indicates a
magnetic field strength of $3.6\times10^{12}$~G, typical of magnetised neutron
stars.

Our VLA radio observations fail to detect the source at a very low level,
indicating that any possible radio emission is at least 60 times weaker than
what would be expected from a black hole system in the low/hard state. This
lack of radio detection is again compatible with the presence of a magnetised
neutron star. \object{4U~2206+54} appears to be the first known system
containing an accreting neutron star that does not show up as an X-ray pulsar,
most likely due to a simple geometrical effect.

Longer high-energy exposures, such as an {\it INTEGRAL} long pointing, are
needed to improve the S/N ratio to a level that will allow the confirmation of
the presence of the cyclotron feature and the study of its eventual profile
and energy changes with time and luminosity. Such observation would also allow
the search for very long (several hours) or very weak pulsations.

\begin{acknowledgements}
We are grateful to the VLA Scheduling Committee, which allowed us to conduct
the observations as an {\it ad hoc} proposal.
We thank Silvia Mart\'{\i}nez N\'u\~nez for very useful discussions about JEM-X data and Elena Gallo for useful clarifications on the X-ray/radio correlation.
We acknowledge useful comments and clarifications from Nicolas Produit and {\it INTEGRAL} Science Data Center members.
We acknowledge an anonymous referee for detailed and useful comments that
helped to improve the paper.
This research is supported by the Spanish Ministerio de Educaci\'on y Ciencia (former Ministerio de Ciencia y Tecnolog\'{\i}a) through grants AYA2001-3092, ESP-2002-04124-C03-02, ESP-2002-04124-C03-03 and AYA2004-07171-C02-01, partially funded by the European Regional Development Fund (ERDF/FEDER). 
P.B. acknowledges support by the Spanish Ministerio de Educaci\'on y Ciencia through grant ESP-2002-04124-C03-02.
M.R. acknowledges support by a Marie Curie Fellowship of the European
Community programme Improving Human Potential under contract number
HPMF-CT-2002-02053. 
I.N. is a researcher of the programme {\em Ram\'on y Cajal}, funded by the Spanish Ministerio de Educaci\'on y Ciencia and the University of Alicante, with partial support from the Generalitat Valenciana and the European Regional Development Fund (ERDF/FEDER).
This research has made use of the NASA's Astrophysics Data System Abstract
Service, and of the SIMBAD database, operated at CDS, Strasbourg, France.
\end{acknowledgements}


\begin{thebibliography}{}

\bibitem[1996]{arnaud96}
Arnaud, K.~A.
1996, in Astronomical Data Analysis Software and Systems V, ASP Conf. Ser., 101, 17

\bibitem[1997]{bildsten97}
Bildsten, L., Chakrabarty, D., Chiu, J., et~al.
1997, ApJS, 113, 367

\bibitem[2001]{clark01}
Clark, J.~S., Reig, P., Goodwin, S.~P., et~al.
2001, A\&A, 376, 476

\bibitem[2002]{clark02}
Clark, J.~S., Goodwin, S.~P., Crowther, P.~A., et~al.
2002, A\&A, 392, 909

\bibitem[2002]{coburn02}
Coburn, W., Heindl, W.~A., Rothschild, R.~E., et~al.
2002, ApJ, 580, 394

\bibitem[1998]{condon98}
Condon, J.~J., Cotton, W.~D., Greisen, E.~W., et~al.
1998, AJ, 115, 1693

\bibitem[2001]{corbet01}
Corbet, R.~H.~D., \& Peele, A.~G.
2001, ApJ, 562, 936

\bibitem[2003]{diehl03}
Diehl, R., Baby, N., Beckmann, V., et~al.
2003, A\&A, 411, L117

\bibitem[2004]{demartino04}
de Martino, D., Matt, G., Belloni, T., Haberl, F., \& Mukai, K.
2004, A\&A, 419, 1009

\bibitem[2000]{fender00}
Fender, R.~P., \& Hendry, M.~A.
2000, MNRAS, 317, 1

\bibitem[1994]{finley94}
Finley, J.~P., Taylor, M., \& Belloni, T.
1994, ApJ, 429, 356

\bibitem[2003]{gallo03}
Gallo, E., Fender, R.~P., \& Pooley, G.~G.
2003, MNRAS, 344, 60

\bibitem[2003]{goldwurm03}
Goldwurm, A., David, P., Foschini, L., et~al.
2003, A\&A, 411, L223

\bibitem[1989]{howarth89}
Howarth, I.~D., \& Prinja, R.~K.
1989, ApJS, 69, 527

\bibitem[1996]{jahoda96}
Jahoda, K., Swank, J.~H., Giles, A.~B., et~al.
1996, in EUV, X-ray, and Gamma-Ray Instrumentation for Astronomy VII, ed. O.~H. Siegmund, \& M.~A. Gummin, SPIE, 2808, 59

\bibitem[2004]{kaper04}
Kaper, L., van der Meer, A., \& Tijani, A.~H.
2004, in Proc. of IAU Colloquium 191, Revista Mexicana de Astronom\'{\i}a
y Astrof\'{\i}sica (Serie de Conferencias), Vol. 21, p. 128

\bibitem[2004]{kreykenbohm04}
Kreykenbohm, I., Wilms, J., Coburn, W., et~al.
2004, A\&A, 427, 975

\bibitem[1988]{leitherer88}
Leitherer, C.
1988, ApJ, 326, 356

\bibitem[1981]{maraschi81}
Maraschi, L., \& Treves, A.
1981, MNRAS, 194, 1P

\bibitem[2005]{martocchia05}
Martocchia, A., Motch, C., \& Negueruela, I.
2005, A\&A, 430, 245

\bibitem[2002]{masetti02}
Masetti, N., Dal Fiume, D., Cusumano, G., et~al.
2002, A\&A, 382, 104

\bibitem[2004]{masetti04}
Masetti, N., Dal Fiume, D., Amati, L, et~al.
2004, A\&A, 423, 311

\bibitem[2004]{massi04}
Massi, M., Rib\'o, M., Paredes, J.~M., et~al.
2004, A\&A, 414, L1

\bibitem[2004]{mcclintock04}
McClintock, J.~E., \& Remillard, R.~A.
2004, in Compact Stellar X-Ray Sources, ed. W.~H.~G. Lewin \& M. van der Klis (Cambridge University Press) in press [{\tt arXiv:astro-ph/0306213}]

\bibitem[2004]{mcswain04}
McSwain, M.~V., Gies, D.~R., Huang, W., et~al.
2004, ApJ, 600, 927

\bibitem[2001]{negueruela01}
Negueruela, I., \& Reig, P.
2001, A\&A, 371, 1056 (NR01)

\bibitem[2004]{negueruela04}
Negueruela, I.
2004, in proceedings of 'The Many Scales of the Universe-JENAM 2004
Astrophysics Reviews, Kluwer Academic Publishers, eds. J. C. del Toro
Iniesta et~al., [{\tt arXiv:astro-ph/0411759}]

\bibitem[2000]{paredes00}
Paredes, J.~M., Mart\'{\i}, J., Rib\'o, M., \& Massi, M.
2000, Science, 288, 2340

\bibitem[2002]{paredes02}
Paredes, J.~M., Rib\'o, M., Ros, E., Mart\'{\i}, J., \& Massi, M.
2002, A\&A, 393, L99

\bibitem[2002]{protassov02}
Protassov, R., van Dyk, D.~A., Connors, A., Kashyap, V.~L., \&
Siemiginowska, A.
2002, ApJ, 571, 545

\bibitem[1996]{reig96}
Reig, P., Chakrabarty, D., Coe, M.~J., et~al.
1996, A\&A, 311, 879

\bibitem[2003]{reig03}
Reig, P., Rib\'o, M., Paredes, J.~M., \& Mart\'{\i}, J.
2003, A\&A, 405, 285

\bibitem[1999]{ribo99}
Rib\'o, M., Reig, P., Mart\'{\i}, J., \& Paredes, J.~M.
1999, A\&A, 347, 518

\bibitem[2005]{ribo05}
Rib\'o, M., Negueruela, I., Torrej\'on, J.~M., Blay, P., \& Reig, P.
2005, A\&A, submitted

\bibitem[1992]{saraswat92}
Saraswat, P., \& Apparao, K.~M.~V.
1992, ApJ, 401, 678 

\bibitem[1999]{shrader99}
Shrader, C.~R. \& Titarchuk, L.
1999, ApJ, 521, L121

\bibitem[2003]{skinner03}
Skinner, G., \& Connell, P.
2003, A\&A, 411, L123

\bibitem[1980]{sunyaev80}
Sunyaev, R.~A., \& Titarchuk, L.~G.
1980, A\&A, 86, 121 

\bibitem[1994]{titarchuk94}
Titarchuk, L.
1994, ApJ, 434, 570

\bibitem[2004]{torrejon04}
Torrej\'on, J.~M., Kreykenbohm, I., Orr, A., Titarchuk, L., \& Negueruela, I.
2004, A\&A, 423, 301

\bibitem[2003]{westergaard03}
Westergaard, N.~J., Kretschmar, P., Oxborrow, C.~A., et~al.
2003, A\&A, 411, L257

\bibitem[1983]{white83}
White, N.~E., Swank, J.~H., \& Holt, S.~S.
1983, ApJ, 270, 711

\bibitem[2003]{winkler03}
Winkler, C., Courvoisier, T.~J.-L., Di Cocco, G., et~al.
2003, A\&A, 411, L1

\end{thebibliography}
\end{document}